\documentclass[conference]{IEEEtran}
\IEEEoverridecommandlockouts
\usepackage{cite}
\usepackage{amsmath,amssymb,amsfonts}
\usepackage{algorithmic}
\usepackage{graphicx}
\usepackage{textcomp}
\usepackage{xcolor}
\def\BibTeX{{\rm B\kern-.05em{\sc i\kern-.025em b}\kern-.08em
    T\kern-.1667em\lower.7ex\hbox{E}\kern-.125emX}}
\begin{document}

\title{A 5G Enabled Adaptive Computing Workflow for Greener Power Grid
\thanks{The Pacific Northwest National Laboratory (PNNL) is operated by Battelle for the U.S. Department of Energy (DOE) under Contract DE-AC05-76RL01830. This work was supported by DOE Office of Science Advanced Scientific Computing Research (ASCR) Program.
 (e-mails: yousu.chen$|$xiaoyuan.fan$|$dexin.wang$|$james.ogle@pnnl.gov, liweiw@g.clemson.edu).
}
}

\author{\IEEEauthorblockN{Yousu Chen$^{1}$, Liwei  Wang$^{2}$, Xiaoyuan Fan$^{1}$, Dexin  Wang$^{1}$, James  Ogle$^{1}$}
\IEEEauthorblockA{$^{1}$Pacific Northwest National Laboratory, Richland, WA, USA \\
$^{2}$Clemson University, North Charleston, SC, USA}
}
\maketitle

\begin{abstract}


5G wireless technology can deliver higher data speeds, ultra low latency, more reliability, massive network capacity, increased availability, and a more uniform user experience to users. It brings additional power to help address the challenges brought by renewable integration and decarbonization. In this paper, a 5G enabled adaptive computing workflow has been presented that consists of various computing resources, such as 5G equipment, edge computing, cluster, Graphics processing unit (GPU) and cloud computing, with two examples showing technical feasibility for edge-grid-cloud interaction for power system real-time monitoring, security assessment, and forecasting. Benefiting from the high speed data transport and massive connection capability of 5G, the workflow shows its potential to seamlessly integrate various applications at distributed and/or centralized locations to build more complex and powerful functions, with better flexibility.

\end{abstract}


\section{Introduction}

Power system is evolving in a fast pace that brings many challenges for grid operation and planning. During this transition, a massive amount of data is generated all over the grid, with new characteristics such as high dynamic, fast response and high uncertainties. People rely on advanced technologies to provide better support in their decision-making process, and help develop solutions keep the lights on. The potential of ultra reliable and low latency communications in 5G hold promise to be an enabler for this transformation, for supporting stringent QoS (Quality of Service) requirements those are imposed to fulfill highly interactive applications, and providing ultra-low latency and high throughput.

5G technologies will allow connecting a larger number of smart grid devices closely to help provide accurate and continuous real-time monitoring, fast control, and timing operation in power industry, based on its supreme computational capability and fast data transfer rate. This is extremely important for the future grid that would involve large-scale distributed energy resources (DERs), Microgrids, and storage, with centralized, distributed, or decentralized structures.

To address the challenges brought by the complex nature of power system transmission and distribution network models, 5G technology could be valuable when exploring a new way to manage grid operation \cite{b1}. There are many researchers recognized the power of 5G. For example, an illustration of 5G digital continuum from sensing and computing perspective is presented in \cite{b2}, Opportunities and challenges for Internet of Thing (IoT)-based 5G network are summarized in \cite{b3}. A remote control application introduced in \cite{b4} shows the effectiveness of combining 5G and image processing at the edge by use of GPU. Several use cases of IoT systems infrastructures empowered by heterogeneous computing architectures are introduced in \cite{b5}. The IoT system with great performance implemented by applying CPU-FPGA heterogeneous computation architecture is presented. Besides, a new hyperconverged platform that supports GPU-based 5G gNB developed by NVIDIA and several use cases discussed in \cite{b6} \cite{b7}. The recent technical report of \cite{b1} discussed the technical characterization and benefit evaluation of 5G enabled grid data transport with some potential applications.


This paper, as a part of effort of an ongoing 5G Energy FRAME project \cite{b8}, is targeted at demonstrating integrated computational workflow based on various computing resources, to explore the technique feasibility for aiding future grid management and control.

This paper is structured as follows. Section II summarizes the advantages brought by 5G to enhance edge computing. Section III describes a 5G enabled integrated workflow environment based on the 5G lab in Pacific Northwest National laboratory (PNNL). Two preliminary examples are presented in Section IV to explore technology feasibility for leveraging 5G in an adaptive computing workflow for better grid management and control, with a discussion on future use cases for the 5G integrated workflow. Section V concludes the paper.
\section{Cloud and 5G Enhanced Edge Computing}
Cloud computing is a comprehensive computational solution integrating servers, storage, network, software, and customized applications. Its architecture can be customized to fit the need of specific application scenarios. Edge computing is a variant topology of distributed computing targeted to location-sensitive applications. With the surge of computation demand on the edge and the emergence of new generation networks, cloud computing can be accommodated to a new paradigm.

\subsection{Edge to Cloud Computing}

In traditional cloud computing scenarios, the data acquired on the user end device, also called user equipment (UE) is uniformly sent to the central cloud server for performing complex computation, data summarization, and secure data storage\cite{b9}. UE is often distributed across a wide range of physical locations to capture macroscopic data samplings. High failure rates and increased energy consumption may be caused by long transmission path. To address the massive data and increased demand for instant response with mobile devices and IoT devices becoming popular, an alternative infrastructure that locates at the edge of the network and migrates the computations closer to the data source is more than needed.

\begin{figure}[htbp]
\centerline{\includegraphics[width=6cm]{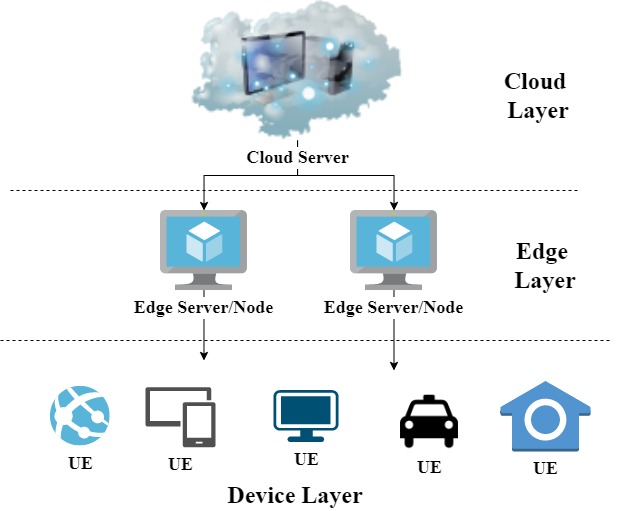}}
\caption{An illustration of Edge Computing Architecture}
\label{fig:architecture}
\end{figure}

The general architecture of edge computing is shown in Fig. \ref{fig:architecture}. It is composed of the cloud layer, edge layer, and device layer. In the device layer, the user end devices take charge of collecting local data and communicating with edge servers or nodes in the edge layer via local area network or wide area network. These edge servers and nodes are responsible for processing and reducing the data collected from the user equipment. Additionally, services like responding to real-time requests, and data visualization are also available for in-situ deployment. The communication between the cloud layer and the edge layer is via the wired or wireless communications. For the cloud layer, the central cloud server takes over the massive data analysis and storage, which requires high computation capability and large-scale storage capabilities.

Edge computing and cloud computing can be regarded as complementary paradigms\cite{b10}. Compared to the user equipment, the edge is equipped with relatively higher computation power and larger storage capability. Edge can perform data processing tasks instead of throwing all the burdens to the cloud server. That helps to reduce the consumption costs of communication, storage, and processing; more importantly, local data processing also better preserves data security and user privacy. Besides, the computation latency can also be reduced. Further, the tasks on the edge will not be interrupted when the network fails. Since the data from the UE has been pre-processed on edge, UE does not need to communicate with the cloud server directly. The edge server will transmit the processed results to the cloud server. Only edge, whose number is much less than UE, establishes connections with the cloud server directly and the network bandwidth connected would be less occupied.

\subsection{5G Enhanced Edge Computing}
5G is the fifth-generation wireless cellular network. Compared to the previous generations, it provides a higher quality network with faster speed, more reliable connections, and more extensive coverage. Three main use cases are supported in 5G New Radio Technologies:
: \begin{itemize}
    \item{Enhanced Mobile Broadband (eMBB): mass deployment of low-powered IoT devices}
    \item{Ultra-reliable Low-Latency Communication(URLLC): extremely reliable low-latency communication}
    \item{Massive Machine-Type Communication (mMTC): deliver high-speed mobile broadband}
\end{itemize}

As for the 5G in edge computing, 5G’s characteristics make it an excellent technology to improve edge computing with enhanced overall network quality. For applications which have a high demand for real-time interaction, 5G technology improves the QoS by reducing the latency and improving reliability. The high upload and download speed contributes to high data transmission rate that leads to increased communication efficiency. Also, running local data processing on the edge avoids establishing direct connections between UEs and the cloud server over the core network. The bandwidth of the core network can be saved.

In addition, due to the large number of distributed UEs cross widespread geological locations, it is challenging to select a proper location for deploying edge servers and to design their coverage scope while balancing the network access requirements and cost. By using 5G, edge servers can be embedded in 5G base stations. These edge servers can easily access the network via 5G signals and do not need the extra construction for setting up the edge server. Also, the signal coverage range of 5G base stations can match the coverage scope of edge servers.

\section{5G Enabled Integrated Workflow Environment}

The high throughput and lower latency of 5G can enable unique streaming analytics and data fusion functions at different levels of the power grid, for various needs from different users and/or stakeholders. Through 5G, diverse data from multiple sources can support more efficient monitoring, control, and coordination amount energy participants in a context of green power grid.

5G with integrated computing could significantly help improve the computational efficiency by balancing computational burden at edge devices, and enhancing data transfer efficiency. This section will describe a general integrated workflow design based on the PNNL 5G Innovation Studio facility that involves various computing resources such as cluster, Graphics processing unit (GPU), cloud and edge computing.

\subsection{PNNL 5G Innovation Studio}
The PNNL Advanced Wireless Communication (AWC) 5G Innovation Studio is the U.S. Department of Energy’s first 5G national laboratory to open a studio powered by Verizon 5G. The 5G lab facility owns application simulators, emulators, and real hardware to send data through the different 5G networks as needed for the specific research question. PNNL has a collection of 5G equipment including 5G NSA (Non-Standalone) with mmWave and 5G SA (Standalone) architectures, spectrum analyzers, and more. It is also equipped with edge and cloud computing resources to support research, testing and evaluation, and partner demonstration. All these elements make it a suitable place for developing and testing the proposed 5g enabled framework with different configurations to support edge-grid-cloud applications \cite{b11}.


\subsection{5G Enabled Adaptive Computing Workflow}

\begin{figure}[htbp]
\centerline{\includegraphics[width=9cm]{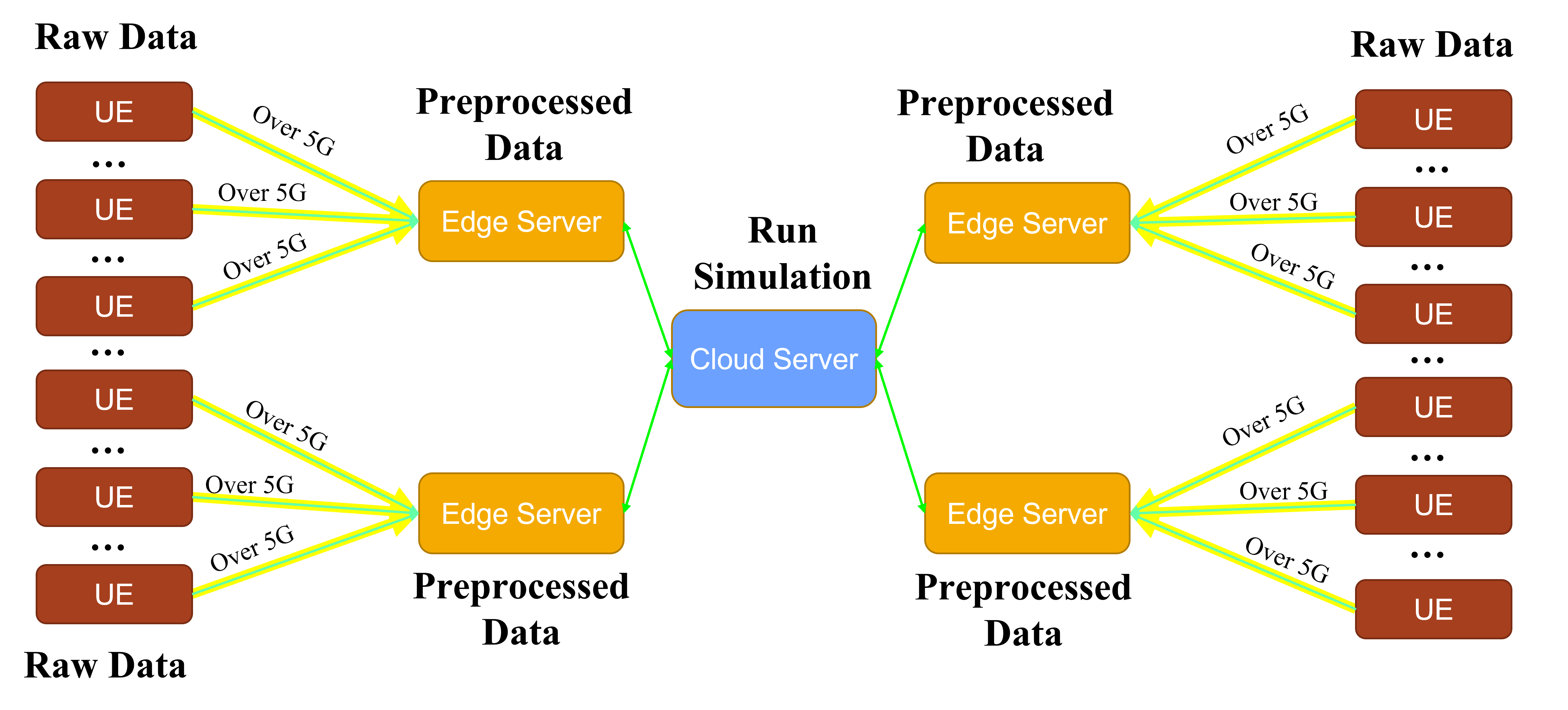}}
\caption{Workflow diagram for the integrated workflow}
\label{fig:generalworkflow}
\end{figure}

The diagram in Fig. \ref{fig:generalworkflow} demonstrates the conceptual workflow of 5G enabled adaptive computing workflow with edge computing connecting to cloud server. This workflow is suitable for the online power system simulation.

Traditionally, the raw data is collected by the user-end data acquisition devices, such as Phasor Measurement Units (PMU) and various smart sensors. These data will be fed into cloud servers, computing machines, and/or high performance computing (HPC) platforms for fast simulation. The simulation results can help power system engineers to make decisions based on the current operating conditions or forecasting values. As shown in Fig. \ref{fig:generalworkflow}, with the new addition of edge server that plays an important role at the intermediate layer for pre-processing the data collected from the user-end device. Each edge server aggregates the data from the neighbor user equipment (UE). The edge server is also equipped with competent computation power so that some lightweight data-processing procedures can be allocated to the edge server.

Involving 5G enhanced edge servers has the following advantages:

\begin{itemize}
    \item{Edge servers are physically closer to the UE, which reduces the data transmission delay and data loss rate. The communication between UE and edge server is via 5G, which could bring better communication efficiency and reliability for transferring the raw data.}
    \item{A large number of UEs can be included in this framework. All UEs need to establish a connection to the central server before they communicate with each other. However, the central server  has limited accommodation for new connections. When the number of requests reaches the upper bound, no new connections are allowed. 5G enhanced edge servers can execute pre-processing tasks that generate intermediate results with lower data volume. Thus, only refined data need to be transmitted to the central server. In this way, the central server can connect to edge servers directly, instead of all the UEs.}
    \item{Real-time decision-making can be achieved with edge. Where there are anomalous behaviors detected, the corrective actions can be directly sent to the responsible UEs to allow timely response.}
\end{itemize}

\subsection{GPU Computing}
 GPU consists of hundreds of processing cores with a large potential to perform large-scale parallel computing. GPU computing leverages these processors to accelerate the computation-intensive scientific problems. To address the growing computation demands on the mobile end, GPU may be deployed widely on the edge servers, besides the dedicated HPC platform or cloud server. The powerful computation capabilities brought by GPU could empower the edge server with more possibilities that more applications can be migrated from the central server to edge, and provide better QoS for the applications with high real-time interaction requirements.

\subsection{5G Experiment Architecture and its Performance}
A 5G-based experimental platform is established to demonstrate the feasibility of the proposed integrated workflow. Three virtual machines are set up with the following roles:
\begin{itemize}
    \item{User Equipment (UE): collect user end data (raw data) and send the user-end data to the edge server over 5G. }
    \item{Edge Server: execute data pre-processing based on the inputs sent from the user equipment and send the processed results to the cloud server.}
    \item{Cloud Server:  run the simulation based on the inputs fed by the edge server. This simulation can take advantages of various computing techniques on the cloud. }
\end{itemize}

\begin{figure}[htbp]
\centerline{\includegraphics[width=7.5cm]{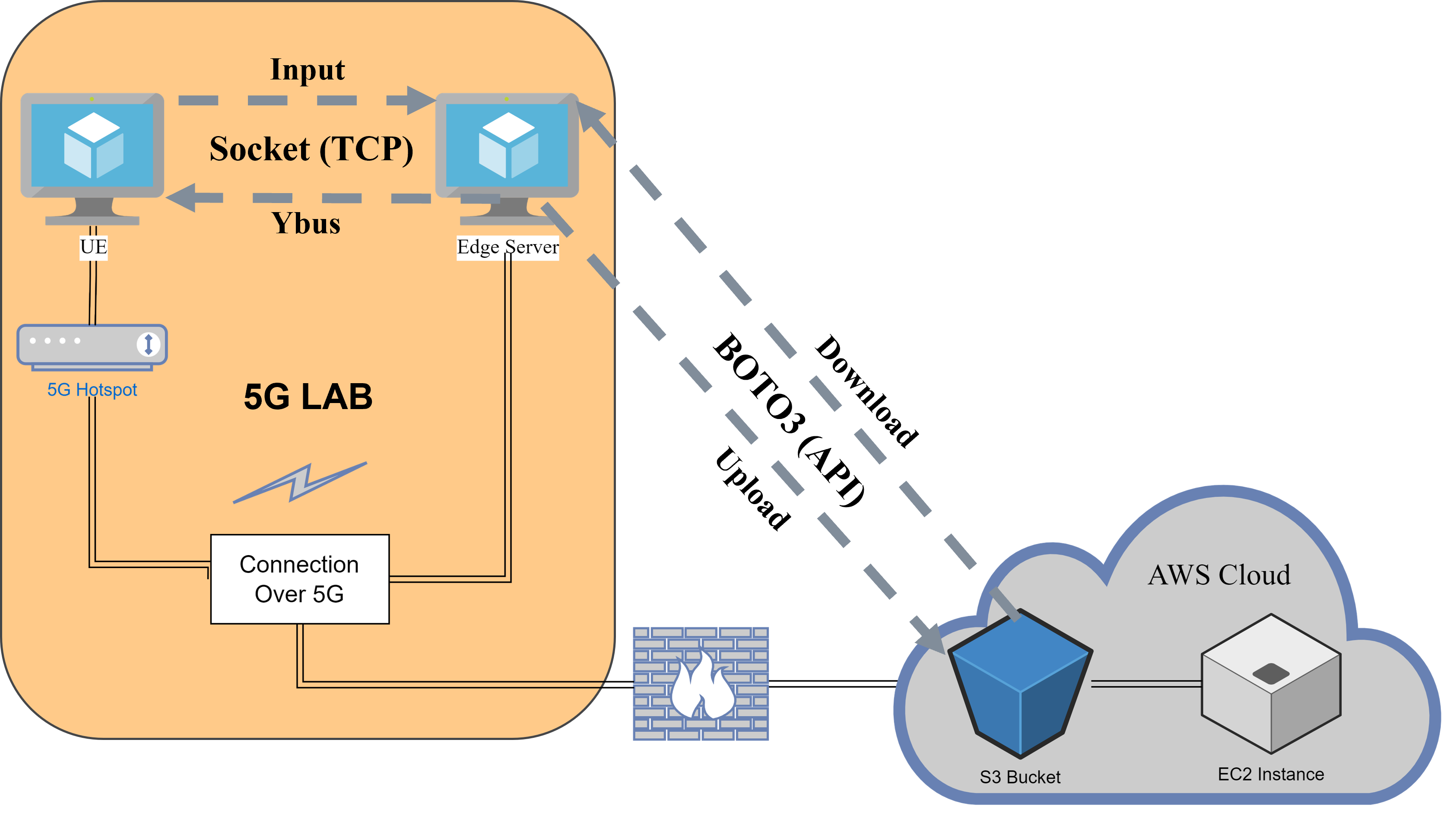}}
\caption{Experiment Platform Architecture}
\label{fig:experimentsetup}
\end{figure}

For this experiment, the data transferred between UE and edge server is based on TCP protocol over 5G, shown in Fig. \ref{fig:experimentsetup}. The facilities within PNNL 5G Innovation Studio supports the standalone  and non-standalone archtectures. All the data sent from the edge server to the cloud server are uploaded into AWS S3 Bucket, a storage solution on the cloud. It also works as intermediate storage media for data transportation. A hybrid CPU+GPU-based dynamic simulation implementation is used in this experiment.

With today's technology, it is possible to send the 5G RF data from a radio directly to an attached computing resource, i.e., GPU, to minimize the latency by performing all the processing of the 5G stack inside the same GPU. This way the GPU becomes an integral part of the 5G devices, with the capability of data processing without incurring the latency for additional data communication with external resources. For example, NVIDIA A100X and A30X \cite{b12} that enable a unified platform of networking and compute are examples for combined capabilities in edge computing and communication.

Early evaluation \cite{b1} on the transmission efficiency of 5G communication has been conducted using an internal LibreSpeed test server \cite{b13} and iperf3\cite{b14} with ping running concurrently. The measurement result of roundtrip latency and roundtrip jitter with the latency of packets show that the average latency is within the range of 15 milliseconds and 37 milliseconds. The corresponding jitter performance spans from 2.5 milliseconds to 18.31 milliseconds. The corresponding average download and upload data transmission rate of 5G Standalone architecture is 306.01 Mbps and 52.43 Mbps, respectively \cite{b1}. These captured performance results are sufficient to support many advanced grid applications.


\section{5G-based adaptive integrated workflow examples}

This paper focuses on the integrated grid simulation platform based on single 5G edge equipment. To show the advantage of 5G edge computing, two use cases have been developed based on the 5G lab hardware testbed and an existing power system parallel dynamic simulation code based on hybrid CPU and GPU structure{\cite{b15}} and smart sampling based scenario selection for dynamic security assessment under uncertainty {\cite{b16}}. The first example is focused on how edge can help send updated distributed information to a centralized location; the second example is about how to leverage forecast capability at edge to help address power system uncertainty quantification problem. Both examples are based on the integrated workflow, with 5G edge, cluster, GPU and cloud computing. Please note the main purpose of these examples are showing a feasible technology path forward, for leveraging 5G technology in helping manage power grids under clear energy transformation; those are also the preludes of a more sophisticate 5G for grid co-simulation use case including transmission, distribution, and communication (T\&D\&C) networks \cite{b1}.

\subsection{5G Enabled Dynamic Simulation with Distributed Online Topology Changes}
\begin{figure}[htbp]
\centerline{\includegraphics[width=8cm]{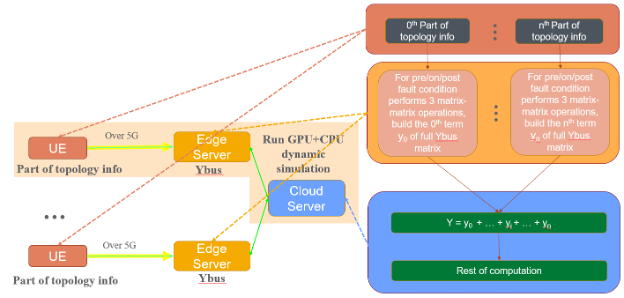}}
\caption{5G Enabled Dynamic Simulation with Distributed Online Topology Changes Workflow}
\label{fig:usecase1}
\end{figure}

The integrated workflow example of dynamic simulation with distributed edge inputs is shown in Fig. \ref{fig:usecase1}. The purpose of this example is to demonstrate the workflow’s capability for online updating the local topology information for dynamic simulation. The hybrid CPU+GPU based dynamic simulation code is used as a base case. The UE is responsible for capturing the local topology information from the user end. The edge server is assigned to calculate the part of the full admittance matrix for pre/on/post fault conditions. In the experiment, a socket server is established on the edge to transfer data between UE and the edge server. When UE captures the local topology data, it will be sent to the edge server immediately. Then the edge starts to execute GPU-based calculation of admittance matrix according to local topology information. When the calculation finishes, the result will be exported and uploaded to the cloud server. When all calculation results from different edge servers have been successfully uploaded to the cloud server, it combines distributed portions of the admittance matrix into the full admittance matrix and runs the dynamic simulation. The simulation results on the cloud will be sent back to the corresponding edge server which contains target topology information for subsequent management and control.

\subsection{5G enabled look-ahead dynamic security assessment under uncertainties}
This example is designed to demonstrate the forecast usability of 5G. Built on top of the first use case, the smart sampling algorithm is integrated into the workflow. We simulate many UEs to send local topology information to edge servers. On the edge, smart sampling\cite{b9} is applied to generate multiple base cases to cover the uncertainty brought by forecast errors. By use of the smart sampling-based probabilistic method and edge computing, the dimension of topology information can be much reduced.

\begin{figure}[htbp]
\centerline{\includegraphics[width=8cm]{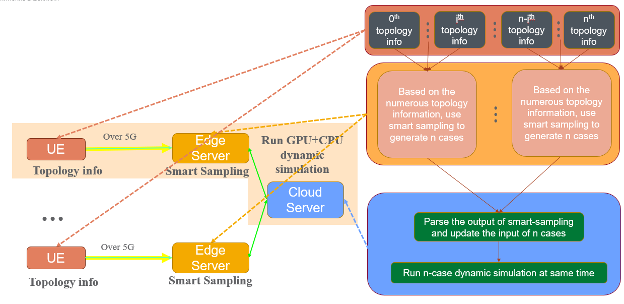}}
\caption{5G enabled look-ahead dynamic security assessment under uncertainty}
\label{fig:usecase2}
\end{figure}

Due to limited number of UE virtual machines, the concurrent data transfer request is simulated in bash scripts. The edge server will apply smart sampling to generate a representative set of scenarios/base cases with the forecast values. When each edge server uploads the smart sampling result to the cloud server, the inputs are updated accordingly. And this set of cases can be run on the cloud server simultaneously for dynamic security assessment under uncertainty. Two-level parallelism with the combination of GPU and CPU is implemented to handle this time-intensive task for optimal speedup performance. This GPU and CPU integration could also potentially help RF procession in 5G.

\subsection{Discussion and future work examples}

\begin{figure}[htbp]
\centerline{\includegraphics[width=8cm]{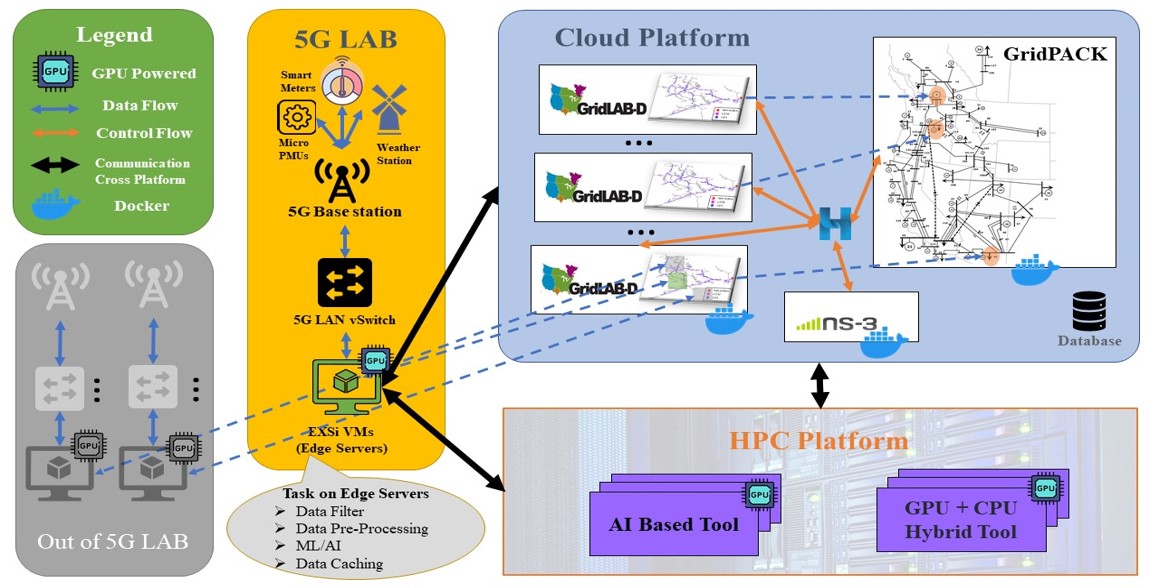}}
\caption{Integrated computation workflow for 5G Energy Frame }
\label{fig:futurework}
\end{figure}

The two examples for the integrated workflow management have clearly showed the technology feasibility to integrate various computing resources and build customized applications based on user needs. In particular, with 5G edge, GPU, cluster and cloud working together, the advantages of each computing resource can be effectively utilized to enable more powerful computation and faster data communication. A high-level example of integrated computational workflow for the 5G based integrated computational workflow is shown in Fig. \ref{fig:futurework}.

The low-latency advantages brought by 5G can help applications that require fast responses, such as fault protection and emergency response, especially for future power electronics dominated grid with DERs and storage. With the help of the proposed 5G enabled adaptive workflow management framework that links to various computing technologies, it can help maximize the benefits during grid evolution stage with effective management and control, such as real-time monitoring, frequency control, hierarchical communications among distributed devices, and system prediction collaboratively while maintaining data privacy by limiting raw data propagation at the edge of the grid. The potential benefits demonstrated by the examples have clearly shown the values, with large spaces of imagining future possibilities.

\section{Conclusion}

This paper briefly summarized the advantages of 5G technology, and presented a 5G enabled adaptive computing integrated workflow that consists of 5G edge computing, cluster, GPU and cloud computing. Two preliminary integrated workflow examples were developed as an initial exploration effort of technology feasibility that could lead to more powerful, more complex power system functions. The proposed methodology and platform is expected to be adapted with different requirements come from the greening power industry and the need to support consumers. Some expected benefits include 5G enabled mobile energy management system for disaster recovery and emergency control with better flexibility and speedy deployment.

\section*{Acknowledgment}

The authors would like to acknowledge the support provided by the PNNL Advanced Wireless Communication (AWC) team and PNNL Center for Advanced Technology Evaluation (CENATE) team.

\vspace{-2ex}

\vspace{12pt}

\end{document}